\documentclass[aps,prx,twocolumn,amsmath,amssymb,longbibliography]{revtex4-2}
\usepackage[latin1]{inputenc}
\usepackage[T1]{fontenc}
\usepackage[english]{babel}
\usepackage[pdftex]{graphicx}
\usepackage{amsmath}
\usepackage{times}
\usepackage[usenames,dvipsnames,svgnames,table]{xcolor}
\usepackage[colorlinks,plainpages=false,linkcolor=blue,urlcolor=blue,citecolor=blue,pdfpagemode=UseNone]{hyperref}
\usepackage{upgreek}
\usepackage{amssymb}
\usepackage{textcomp}
\usepackage{ragged2e}

\renewcommand{\AA}{\text{\r{A}}}

\begin{document}

\title{\boldmath
Tuning of the carrier localization, magnetic and thermoelectric properties in ultrathin (LaNiO$_{3-\delta}$)$_1$/(LaAlO$_{3}$)$_1$(001) superlattices by oxygen vacancies}

\author{Manish Verma}
\email{manish.verma@uni-due.de}
\affiliation{Department of Physics and Center for Nanointegration (CENIDE), Universit\"at Duisburg-Essen, Lotharstr.~1, 47057 Duisburg, Germany}
\author{Rossitza Pentcheva}
\email{rossitza.pentcheva@uni-due.de}
\affiliation{Department of Physics and Center for Nanointegration (CENIDE), Universit\"at Duisburg-Essen, Lotharstr.~1, 47057 Duisburg, Germany}
\date{\today}

\begin{abstract}
Using a combination of density functional theory calculations with an on-site Coulomb repulsion term (DFT+$U$) and Boltzmann transport theory within the constant relaxation time approximation, we explore the effect of oxygen vacancies on the electronic, magnetic, and thermoelectric properties in ultrathin (LaNiO$_{3-\delta}$)$_1$/(LaAlO$_{3}$)$_1$(001) superlattices (SLs). For the pristine SL,  an antiferromagnetic charge-disproportionated (AFM-CD) ($d^{8}${$\underline L$}$^{2}$)$_{S=0}$($d^{8}$)$_{S=1}$ phase is stabilized, irrespective of strain. At $\delta$ = 0.125 and 0.25, the localization of electrons released from the oxygen defects in the NiO$_{2}$ plane triggers a charge-disproportionation, leading to a ferrimagnetic insulator both at $a_{\mathrm{STO}}$ (tensile strain) and $a_{\mathrm{LSAO}}$ (compressive strain). At $\delta$ = 0.5, an insulating phase emerges with alternating stripes of Ni$^{2+}$ (high-spin) and Ni$^{2+}$ (low-spin) and oxygen vacancies ordered along the [110] direction (S-AFM), irrespective of strain. This results in a robust $n$-type in-plane power factor of 24~$\mu$W/K$^2$ cm at $a_{\mathrm{STO}}$ and 14~$\mu$W/K$^2$ cm at $a_{\mathrm{LSAO}}$ at 300~K (assuming relaxation time $\tau = 4$~fs). Additionally, the pristine and $\delta$ = 0.5 SLs are shown to be dynamically stable. This demonstrates the fine tunability of electronic, magnetic, and thermoelectric properties of ultrathin nickelate superlattices by oxygen vacancies.
\end{abstract}


\maketitle

\section{Introduction}\label{SEC1}
Understanding the physics of strongly correlated electrons in transition metal oxides (TMO) in the ultrathin limit is at the forefront of condensed matter physics, as they  are susceptible to collective ordering phenomena resulting in spin- and charge-ordered phases different from the bulk~\cite{Zubko2011, Lorenz2016, Ramesh2019, Mannhart1607, Ali2022}. Engineering such phases became viable due to the possibility to grow such heterostructures with atomically sharp interfaces ~\cite{Mannhart1607}. 

In the context of designing novel quantum phases, bulk LaNiO$_{3}$ -- a paramagnetic metal down to the lowest temperature ($T$)~\cite{Medarde_1997} represents a model system. In contrast, the other members of the rare-earth-nickelate series ($R$NiO$_3$, $R$ = La-Lu), by virtue of small or negative charge transfer energy~\cite{Mizokawa2000, PARK2012, LAU2013, Mizokawa1995, Bodenthin_2011, Abbate2002, Horiba2007, HAN2012, Anisimov1999}, undergo a metal-to-insulator transition (MIT) due to charge-disproportionation (CD)~\cite{PARK2012, GREEN2016, Jhonston2014, Bisogni2016}. This is further accompanied by a symmetry lowering from an orthorhombic ($Pbnm$) to a monoclinic ($P2_1/n$) structure at low $T$, and additionally with an onset of antiferromagnetic (AFM) order in the case of PrNiO$_3$ and NdNiO$_3$~\cite{Torrance8209}. In this picture, MIT stems from the volume collapse of half of the NiO$_6$ octahedra with two ligand holes around the central Ni ($d^{8}${$\underline L$}$^{2}$) site, while the other octahedra expand accordingly with little effect on the net volume of the unit cell, as observed experimentally~\cite{Medarde_1997}. This is viewed as ($d^{8}${$\underline L$}) ($d^{8}${$\underline L$}) $\rightarrow$ ($d^{8}${$\underline L$}$^{2}$)$_{S=0}$($d^{8}$)$_{S=1}$, where $S$ is the total spin, spanning both Ni and O sites. Moreover, the  $d^{8}${$\underline L$}$^{2}$ of the collapsed octahedra is  equivalent to a 3$d^{6}$ occupation that is low spin. Notably, in the case of charge-disproportionation or breathing-mode distortion, the two Ni sites (($d^{8}${$\underline L$}$^{2}$)$_{S=0}$ and ($d^{8}$)$_{S=1}$) remain Jahn-Teller inactive, for instance in LuNiO$_{3}$~\cite{Mazin2007}.

In contrast to bulk, when LaNiO$_3$ is grown as an ultrathin superlattice (SL), it undergoes a MIT below a critical thickness of two monolayers (ML), triggered by the effect of confinement~\cite{Freeland2011, Blanca2011, Geisler2018, Scherwitzl2009, King2014, Kumah2014, Gabay2014, SON2010, Scherwitzl2011, Sakai2013, Liu2011}. Apart from MIT, an antiferromagnetic charge-disproportionated (AFM-CD) phase was observed in (LaNiO$_{3}$)$_{2}$/(LaAlO$_{3}$)$_{2}$(001) SLs upon cooling, irrespective of strain~\cite{Boris937}. Furthermore, Middey $et~al.$~\cite{Middey2018} also recently reported an AFM-CD phase leading to MIT in ultrathin $R$NiO$_3$ SLs. Previous DFT + $U$ studies~\cite{Freeland2011, Blanca2011, Geisler2018} identified charge-disproportionation on the Ni sites with ferromagnetic coupling (FM-CD) as a possible mechanism behind the MIT in (LaNiO$_{3}$)$_{1}$/(LaAlO$_{3}$)$_{1}$(001) (1/1) SL at the lateral lattice constant of  SrTiO$_{3}$ (tensile strain), and a semi-metallic phase with less pronounced CD at compressive strain. On the other hand, Puggioni $et~al.$~\cite{Puggioni2012} reported an AFM-CD phase in (LaNiO$_{3}$)$_{1}$/(LaAlO$_{3}$)$_{1}$(001) SL using a laterally elongated supercell and the variational self-interaction-corrected local density functional theory~\cite{Filippetti2009, Archer2011}, irrespective of strain. They argued that the DFT + $U$ approach is insufficient to describe their findings. Hence the question arises whether DFT + $U$ can explain the AFM-CD insulating phase in (LaNiO$_{3}$)$_{1}$/(LaAlO$_{3}$)$_{1}$(001) SL. 

Apart from LaNiO$_{3}$ bulk and superlattices, there is a recent interest in understanding the electronic and magnetic properties of oxygen-deficient LaNiO$_{3-\delta}$ bulk~\cite{Wang2018, DEY2019, Guo2018, Liu2020, Liao2021} and thin films~\cite{Kawai2009, Tung2017}. Oxygen vacancies are also present in other bulk $R$NiO$_{3}$. Recently, Kotiuga \textit{et al.}~\cite{Kotiuga21992, Kotiuga2019} reported that the electrons released by oxygen vacancies in bulk SmNiO$_3$ and NdNiO$_3$ were localized at the Ni sites, opening a Mott insulating gap, stemming from the interplay of crystal field splitting (CFS) and strongly correlated electrons in the Ni-3$d$ orbitals. Using DFT + $U$, Shin $et~al.$ explored the electronic and magnetic structure of bulk LaNiO$_{3-\delta}$ ($\delta$ = 0.25 and 0.5)  with ordered oxygen vacancies~\cite{Shin2022}. Furthermore, although oxygen vacancies create localized defect states below the Fermi level in bulk LaNiO$_{3}$, a gap is not formed, at least at low oxygen vacancy concentrations~\cite{Malashevich2015}. 

Oxygen defects are ubiquitous also in ultrathin LaNiO$_{3}$/LaAlO$_3$(001) SLs, either under tensile strain~\cite{Boris937} or due to the polar interface to the SrTiO$_3$ substrate~\cite{Liu2010}. Hence it is important to clarify, how oxygen vacancies influence the electronic and magnetic properties  of a (LaNiO$_3$)$_1$/(LaAlO$_3$)$_1$(001) SL and whether they can cause a charge localization induced gap opening  under confinement and strain. While a recent DFT + $U$ study investigated the electronic and magnetic properties of the oxygen deficient (LaNiO$_{3}$)$_{1}$/(LaAlO$_{3}$)$_{1}$(001) SLs, at low oxygen vacancy concentrations~\cite{Geisler2022}, here we concentrate on (LaNiO$_{3-\delta}$)$_{1}$/(LaAlO$_{3}$)$_{1}$(001) SLs with $\delta=0.125-0.50$. 

Furthermore, Yu $et~al.$~\cite{Yu2008} recently reported that the introduction of a moderate oxygen vacancy concentration reduces the lattice thermal conductivity in SrTiO$_{3}$~\cite{Yu2008}. Thus we can expect possibly enhanced scattering of phonons in the oxygen-deficient nickelate SLs which is crucial in optimizing the overall figure of merit ($ZT$). Hence exploring the thermoelectric response in (LaNiO$_{3-\delta}$)$_1$/(LaAlO$_{3}$)$_1$(001) SLs might provide useful insights in this direction.

To address these questions, in this work using DFT + $U$ calculations we explore the electronic and magnetic properties of (LaNiO$_{3-\delta}$)$_1$/(LaAlO$_{3}$)$_1$(001) SL ($\delta=0.0-0.50$). We start with the electronic and structural properties of bulk LaNiO$_{3}$ to ensure that the confined monolayer of LaNiO$_{3}$ in (LaNiO$_{3}$)$_1$/(LaAlO$_{3}$)$_1$(001) SL is under tensile strain at $a_{\mathrm{STO}}$ and compressive strain at $a_{\mathrm{LSAO}}$. Next we turn to the pristine (LaNiO$_{3}$)$_1$/(LaAlO$_{3}$)$_1$(001) SL. By using a larger lateral unit cell~\cite{Verma2022} we find the stabilization of an AFM-CD ($d^{8}${$\underline L$}$^{2}$)$_{S=0}$($d^{8}$)$_{S=1}$ phase and compare this with the metastable FM-CD phase at tensile ($a_{\mathrm{STO}}$) and compressive ($a_{\mathrm{LSAO}}$) strain, respectively. In addition, we obtain the lattice dynamic stability of the pristine (LaNiO$_{3}$)$_1$/(LaAlO$_{3}$)$_1$(001) SL. 

Furthermore, by varying the oxygen vacancy concentration in (LaNiO$_{3-\delta}$)$_1$/(LaAlO$_{3}$)$_1$(001) SL, we show that charge localization in conjunction with confinement and strain triggers an electronic reconstruction leading to a ferrimagnetic ordering at $\delta$ = 0.125 and 0.25, respectively, irrespective of strain. Moreover, at $\delta$ = 0.5, an insulating phase emerges with alternating stripes of Ni$^{2+}$ (high-spin) and Ni$^{2+}$ (low-spin) and oxygen vacancies ordered along the [110] direction at tensile and compressive strain, respectively. We also prove the lattice dynamic stability of this phase. Overall we find that the confinement induced MIT in (LaNiO$_{3-\delta}$)$_1$/(LaAlO$_{3}$)$_1$(001) SLs occurs independent of the oxygen vacancy concentration and strain. 

Lastly, using Boltzmann transport theory within the constant relaxation time approximation, we show how oxygen vacancies can be used to effectively tune the localization of Ni 3$d$ states in ultrathin nickelate superlattices, subsequently impacting the thermoelectric response. At $\delta$ = 0.5, a combination of flat and dispersive bands in the vicinity of the conduction band edge result in simultaneously high values of both the Seebeck coefficient and the electrical conductivity, respectively, together leading to a promising in-plane $n$-type power factor of 24~$\mu$W/$K^2$ cm at $a_{\mathrm{STO}}$ and 14~$\mu$W/$K^2$ cm at $a_{\mathrm{LSAO}}$, respectively, (assuming $\tau$ = 4 fs) at 300~K. This can be compared with some of the best performing oxide thermoelectrics, such as La- or Nb-doped SrTiO$_3$~\cite{Jalan2010, Okuda2001, Ohta2005} or recent predictions for Sr$X$O$_3$/SrTiO$_3$(001) SLs, $X$ = V, Cr, or Mn~\cite{Manish2019, Verma2022}. 

\section{Computational details}\label{SEC2} 
First principles calculations for LaNiO$_{3-\delta}$ and (LaNiO$_{3-\delta}$)$_1$/(LaAlO$_{3}$)$_1$(001) SLs ($\delta$ = 0, 0.125, 0.25 and 0.5) were performed within the framework of spin-polarized DFT~\cite{Kohn1965} using the Vienna $ab$ $initio$ simulation package ($\mathrm{VASP}$)~\cite{Kresse1993, Kresse1996} with the projector augmented wave (PAW) basis~\cite{Bloch1994, Kresse1999}. The generalized gradient approximation was used for the exchange correlation functional in the implementation of Perdew, Burke and Ernzerhof~\cite{Perdew1996}. In order to take static correlation effects into account, the DFT + $U$ approach~\cite{Anisimov1993} with $U$ = 4 eV for the Ni 3$d$ and $U$ = 7 eV for the La 4$f$ was employed within Dudarev's scheme~\cite{Dudarev1998}.

We model the 1/1 SLs by using laterally enlarged 2$\sqrt{2}a \times 2\sqrt{2}a \times 2c$ supercells that contain 80 atoms in total~\cite{Verma2022}. To model epitaxial growth under tensile and compressive strain, we fixed the in-plane lattice constant to the experimental lattice constants of SrTiO$_3$ ($a_{\mathrm{STO}}$ = 3.905~{\AA}) and LaSrAlO$_4$ ($a_{\mathrm{LSAO}}$ = 3.756~{\AA}) reported in Ref.~\cite{Boris937}, respectively, whereas the $c$ parameter along with the internal positions were fully optimized until the forces were less than 0.001 eV/{\AA}. To sample the Brillouin zone, we have used 5$\times$5$\times$6 Monkhorst-Pack $k$-points~\cite{Monkhorst1976} for the $1/1$ superlattices.

To confirm that the confined monolayer of LaNiO$_{3}$ in (LaNiO$_{3}$)$_1$/(LaAlO$_{3}$)$_1$(001) SL is under tensile strain at $a_{\mathrm{STO}}$ and compressive strain at $a_{\mathrm{LSAO}}$, a non-spin polarized calculation was performed on bulk LaNiO$_{3}$ with PBEsol + $U$ to simultaneously optimize the lattice parameters and the internal coordinates (volume-cell relaxation). The rhombohedral cell (R$\bar{3}$c, space-group:167) in the hexagonal set up~\cite{Torrance1992} was adopted, containing 30 atoms. A $\Gamma$ centered $k$-mesh of 9$\times$9$\times$3 was used. In addition, the effect of the Hubbard $U$ term on the electronic and magnetic properties of (LaNiO$_{3}$)$_1$/(LaAlO$_{3}$)$_1$(001) SLs was studied by  optimizing the volume with a varying value of $U_{\rm Ni}$ from 1 to 5 eV in spin-polarized calculations. The role of La 4$f$ states was assessed by carrying out calculations with $U_{\rm La}$ = 0 and 7 eV.

To prove the dynamic stability of the ground state SL structures and magnetic orders in (LaNiO$_{3}$)$_1$/(LaAlO$_{3}$)$_1$(001) and (LaNiO$_{2.5}$)$_1$/(LaAlO$_{3}$)$_1$(001) SLs, respectively, lattice dynamics calculations were performed on the volume-cell relaxed structures (obtained with PBEsol + $U$, $U_{\mathrm{Ni}}$ = 4 eV and $U_{\mathrm{La}}$ = 7 eV) using the supercell finite-differences approach implemented in the Phonopy package~\cite{Togo2023A, Togo2023B}. The second-order (harmonic) force constants were calculated using 1$\times$1$\times$1 and 1$\times$1$\times$2 supercells of the (LaNiO$_{3}$)$_1$/(LaAlO$_{3}$)$_1$(001) SL and the (LaNiO$_{2.5}$)$_1$/(LaAlO$_{3}$)$_1$(001) SL leading to 60 and 116 displacements, respectively. Note that the 1$\times$1$\times$2 supercells are w.r.t. the above mentioned 2$\sqrt{2}a \times 2\sqrt{2}a \times 2c$ system. Atom-projected vibrational density of states curves were computed by interpolating the phonon frequencies on to a regular $\Gamma$ centered $q$-point 15$\times$15$\times$18 grid using the linear tetrahedron method for Brillouin zone integration. Phonon dispersions were obtained by evaluating frequencies at strings of $q$ points along the high-symmetry points in the respective Brillouin zone by employing the Sumo package~\cite{Alex2018}.

Lastly, we have calculated the thermolectric properties employing Boltzmann transport theory in the constant relaxation time approximation, as implemented in the BoltzTraP code~\cite{MADSEN2006} using a very dense $k$-mesh of $20\times20\times24$. For calculating the thermoelectric quantities we have utilized the approach of Sivan and Imry~\cite{Sivan1986}, described and used in previous studies~\cite{Geisler2017, Geisler2018, Geisler2019}.

\begin{figure}
	\includegraphics[width=8.8cm, height=4.3cm]{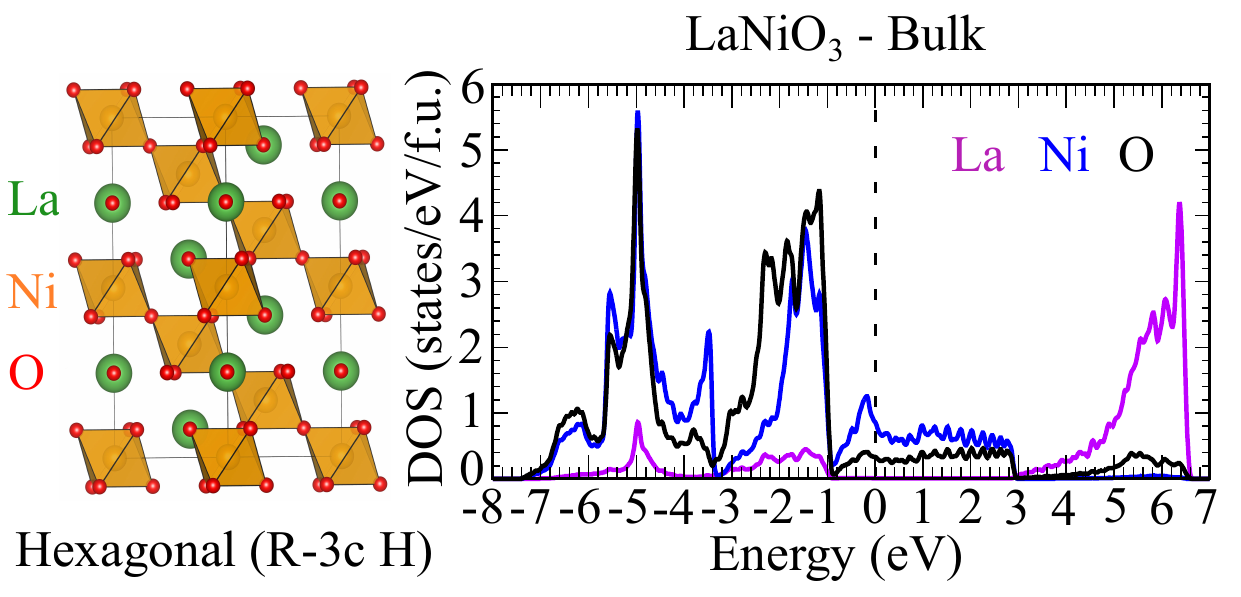}
	\caption{\label{Fig:HS} Left: Side view of the volume-cell relaxed hexagonal cell of bulk LaNiO$_{3}$. Right: Element resolved density of states (DOS) of bulk LaNiO$_{3}$. Magenta, blue and black color represents the La, Ni and O states, respectively.}
	\label{Fig1}
\end{figure}

\begin{figure*}
	\centering
	\includegraphics[width=16.5cm, height=9.3cm]{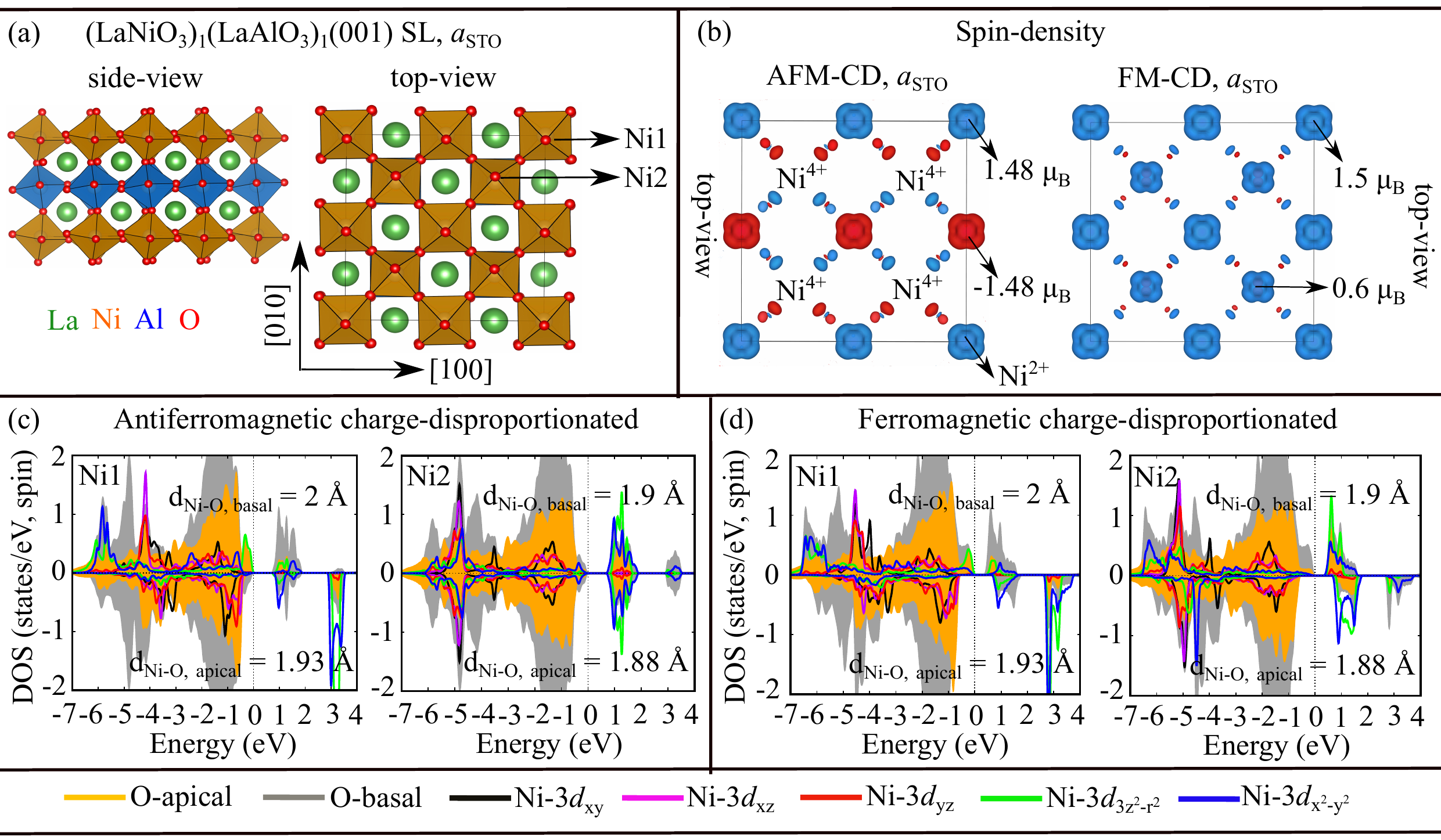}
	\caption{\label{Fig:HS}(a) Side and top views of the optimized pristine (LaNiO$_{3}$)$_1$/(LaAlO$_{3}$)$_1$(001) SL at $a_{\mathrm{STO}}$, (b) Spin densities (side and top views), integrated from -7~eV to the Fermi energy (0~eV), together with the local magnetic moments at the two distinct Ni sites. Blue and red colors correspond to positive and negative spin densities, respectively. (c), (d) Orbital-projected DOS of the two distinct Ni sites having antiferromagnetic and ferromagnetic charge-disproportionation, respectively, along with the density of states of the apical and basal oxygen associated with their corresponding octahedra. The isovalue of the spin-density is 0.006 e/\AA$^{3}$.}
	\label{Fig2}
\end{figure*}

\section{\boldmath Electronic and structural properties of bulk LaN\lowercase{i}O$_{3}$}\label{SEC3}

We start by discussing the electronic and structural properties of bulk LaNiO$_{3}$. From the volume-cell relaxed hexagonal structure of bulk LaNiO$_{3}$ (shown in Fig.~\ref{Fig1}, left), we obtained the pseudo-cubic lattice constant of 3.80~{\AA} which is in excellent agreement with the experimental value of 3.832~{\AA} at $T$ = 1.5 K~\cite{Torrance1992}. Noting the experimental lattice constants of bulk LaAlO$_{3}$ (3.792~{\AA}), SrTiO$_{3}$ (3.905~{\AA}), and LSAO (3.756 ~{\AA}), it is confirmed that the confined monolayer of LaNiO$_{3}$ in (LaNiO$_{3}$)$_1$/(LaAlO$_{3}$)$_1$(001) SL is under tensile strain at $a_{\mathrm{STO}}$ and compressive strain at $a_{\mathrm{LSAO}}$, respectively. The element-resolved density of states (DOS) in Fig.~\ref{Fig1} (right), exhibits metallic behavior in agreement with the experimental findings~\cite{Medarde_1997}. The states between -1 to 3 eV are predominantly of Ni-3$d$ character with lower contribution from O states, whereas  Ni 3$d$ states with prominent peaks, admixed with comparable O 2$p$ states can be observed between -7 to -1 eV, indicating strong $p$-$d$ hybridization. The La-4$f$ states remain localized between 5 to 7 eV with $U_{\mathrm{La}}$ = 7 eV.

\section{\boldmath Electronic and magnetic properties of the pristine (L\lowercase{a}N\lowercase{i}O$_{3}$)$_1$/(L\lowercase{a}A\lowercase{l}O$_{3}$)$_1$(001) superlattice}\label{SEC4}

To understand the interplay of confinement and strain, we now discuss the electronic and magnetic properties of the pristine SL. The optimized cross-plane lattice constants are 3.74~{\AA} and 3.88~{\AA} at tensile and compressive strain, respectively. The partial occupation of the degenerate $e_g$ orbitals is accommodated by charge disproportionation ($d^{8}${$\underline L$}) ($d^{8}${$\underline L$}) $\rightarrow$ ($d^{8}${$\underline L$}$^{2}$)$_{S=0}$($d^{8}$)$_{S=1}$, including both Ni and O in the NiO$_6$ octahedra, irrespective of strain. Similar to the low-temperature phase of $R$NiO$_{3}$~\cite{Jhonston2014}, the pristine (LaNiO$_{3}$)$_1$/(LaAlO$_{3}$)$_1$(001) SL exhibits a \textgravedbl breathing-mode\textacutedbl ~distortion, which creates expanded ($S$ = 1) and compressed ($S$ = 0) NiO$_{6}$ octahedra with volume of 10.61 and 9 {\AA$^3$} ($a_{\mathrm{STO}}$) and 10 and 8.63 {\AA$^3$} ($a_{\mathrm{LSAO}}$) at the Ni1 and Ni2 sites (Fig.~\ref{Fig2} (a)), respectively, forming a two-dimensional checkerboard pattern. Notably, the Ni moments are arranged in a $\cdots$ $\uparrow$ $\cdot$ 0 $\cdot$ $\downarrow$ $\cdot$ 0 $\cdots$ pattern along the [100] direction (see Fig.~\ref{Fig2} (b)-left). Such a confinement-induced electronic reconstruction leads to the opening of a band gap which is smaller under tensile strain (0.79 eV) than compressive strain (1 eV), along with a slightly enhanced Ni$^{2+}$ spin magnetic moment ($1.48~\mu_\text{B}$) at tensile strain, compared to ($1.44~\mu_\text{B}$) at compressive strain. 

At $a_{\mathrm{STO}}$/$a_{\mathrm{LSAO}}$, the AFM-CD ($d^{8}${$\underline L$}$^{2}$)$_{S=0}$($d^{8}$)$_{S=1}$ phase was found to be more stable than the FM-CD~\cite{Blanca2011}, stripe AFM-CD (S-AFM-CD), and G-AFM magnetic order by 17/11 meV/Ni atom, 49/24 meV/Ni atom, and 81/131 meV/Ni atom, respectively. Note that the S-AFM-CD magnetic order comprises of alternating stripes of Ni moments having $\cdots$ $\uparrow$ $\cdot$ 0 $\cdot$ $\uparrow$ $\cdot$ 0 $\cdots$ and $\cdots$ $\downarrow$ $\cdot$ 0 $\cdot$ $\downarrow$ $\cdot$ 0 $\cdots$ pattern, respectively, along the [$\Bar{1}$10] direction. We compare the electronic and magnetic properties of the ground state AFM-CD phase with the metastable phases at both tensile and compressive strain, respectively. By analyzing the orbital-projected DOS for the AFM-CD phase in Fig.~\ref{Fig2} (c) for $a_{\mathrm{STO}}$ (see Fig. S1-(c) of the supplemental material~\cite{Appendix2020} for $a_{\mathrm{LSAO}}$), we observe that both $e_g$ orbitals are nearly singly occupied and hybridized with apical oxygen states (orange filled curves), along with fully filled $t_{2g}$ orbitals ($d^{8}$) below the Fermi level at the Ni1O$_{6}$ octahedra. Furthermore, the basal oxygen is slightly spin-polarized ($\sim$ 0.03~$\mu_\text{B}$), followed by a negligible spin-polarization and a suppression of holes at the apical oxygens. On the other hand, in the Ni2O$_6$ octahedra only the $t_{2g}$ orbitals are filled in both spin channels, leading to a zero spin moment, along with predominantly depleted apical oxygen states (orange filled curves) below the Fermi level ($d^{8}${$\underline L$}$^{2}$). Overall the confinement leads to an AFM-CD ($d^{8}${$\underline L$}$^{2}$)$_{S=0}$($d^{8}$)$_{S=1}$ phase, irrespective of strain, similar to the bulk rare-earth nickelates~\cite{Jhonston2014}. In the FM-CD phase, the charge disproportionation is somewhat less pronounced, reflected in distinct magnetic moments of $1.5/0.6~\mu_\text{B}$ ($a_{\mathrm{STO}}$) and $1.43/0.54~\mu_\text{B}$ ($a_{\mathrm{LSAO}}$) at the Ni1/Ni2 sites, shown in the spin density plot in Fig.~\ref{Fig2} and Fig. S1 of the supplemental material~\cite{Appendix2020}, respectively; consistent with previous studies~\cite{Blanca2011, Geisler2018}. We also find distinct breathing-mode distortions and magnetic moments for S-AFM-CD order: $0/1.55~\mu_\text{B}$ ($a_{\mathrm{STO}}$) and $0/1.5~\mu_\text{B}$ ($a_{\mathrm{LSAO}}$) with the corresponding octahedral volumes 9/10.7 {\AA$^3$} ($a_{\mathrm{STO}}$) and 8.64/10.13 {\AA$^3$} ($a_{\mathrm{LSAO}}$) at the Ni1/Ni2 sites, respectively. Interestingly, a similar magnetic ordering dependent breathing-mode distortion has been reported previously in bulk NdNiO$_{3}$~\cite{Stoica2022}. On the other hand, for G-AFM order, a uniform magnetic moment of $0.7~\mu_\text{B}$ was obtained at all  Ni sites, irrespective of strain. This goes hand-in-hand with a depletion of states below the Fermi level of the apical O-2$p$ at all the octahedra, irrespective of strain (see Fig. S2 of the supplemental material~\cite{Appendix2020}).
Our results highlight that the magnetic order and strain have direct impact on the degree of breathing-mode distortion in the pristine (LaNiO$_{3}$)$_1$/(LaAlO$_{3}$)$_1$(001) SL. Finally, we have also studied the effect of Hubbard $U$ at Ni-3$d$ states on the electronic and magnetic properties of the pristine SLs (see Sec. S-2 of the supplemental material~\cite{Appendix2020}). We adopted $U_{\mathrm{Ni}}$ = 4 eV in this work as it renders the charge-disproportionation in agreement with previous studies~\cite{Blanca2011, Geisler2018}.

\begin{figure*}
	\includegraphics[width=18.1cm, height=6.3cm]{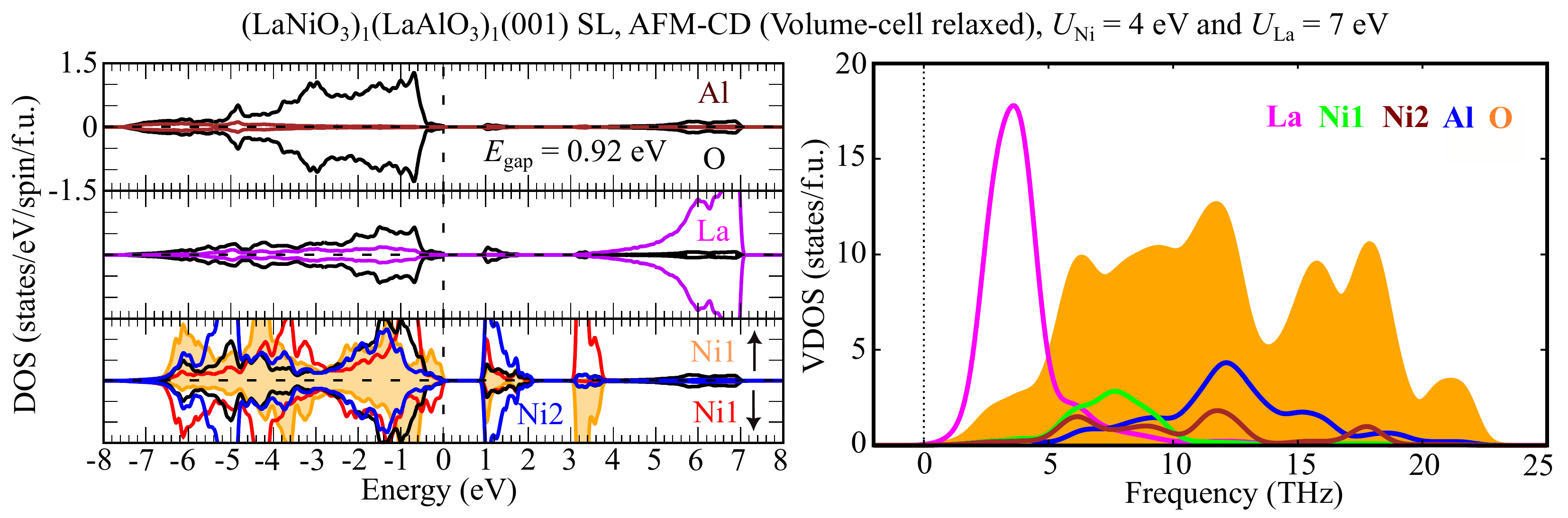}
	\caption{\label{Fig:HS} Layer-, element-, and spin-resolved DOS (left panel) and corresponding atoms projected vibrational density of states curves (right panel) of the volume-cell relaxed (LaNiO$_{3}$)$_1$/(LaAlO$_{3}$)$_1$(001) SL with AFM-CD (($d^{8}${$\underline L$}$^{2}$)$_{S=0}$($d^{8}$)$_{S=1}$) magnetic order, obtained using $U_{\mathrm{Ni}}$ = 4 eV and $U_{\mathrm{La}}$ = 7 eV, respectively. In the projected density of states, magenta, black and brown color represent La, O, and Al states, respectively. Additionally in the NiO$_{2}$ panel, filled orange and red line curves represent the spin-up and spin-down states, respectively, of the Ni1 (($d^{8}$)$_{S=1}$) atom, whereas, blue line curve represents the spin-polarized states of the Ni2 (($d^{8}${$\underline L$}$^{2}$)$_{S=0}$) atom. The dashed line at the zero energy represents the Fermi level. In the vibrational density of states, magenta, green, brown, blue and orange color represent La, Ni1, Ni2, Al, and O states, respectively.}
	\label{Fig3}
\end{figure*}

\section{\boldmath Lattice dynamical stability of pristine (L\lowercase{a}N\lowercase{i}O$_{3}$)$_1$/(L\lowercase{a}A\lowercase{l}O$_{3}$)$_1$(001) superlattice}\label{SEC5}

In this section we turn to the lattice dynamical stability of (LaNiO$_{3}$)$_1$/(LaAlO$_{3}$)$_1$(001) SL in the AFM-CD (($d^{8}${$\underline L$}$^{2}$)$_{S=0}$($d^{8}$)$_{S=1}$) phase. In contrast to the strained SLs considered in section~\ref{SEC4}, here we have adopted the volume-cell relaxed SL structure, obtained by employing $U_{\mathrm{Ni}}$ = 4 eV and $U_{\mathrm{La}}$ = 7 eV, respectively. The optimized in-plane and cross-plane lattice parameters are 3.81~{\AA} and 3.8~{\AA}, respectively. Using the laterally enlarged 2$\sqrt{2}a \times 2\sqrt{2}a \times 2c$ cell in the AFM-CD phase, the symmetry reduced spontaneously from the monoclinic ($P2_{1}/c$) to triclinic ($P1$) symmetry, similar to the strained SLs, considered in section~\ref{SEC4}. From the layer-, element- and spin-resolved DOS in Fig.~\ref{Fig3} (left), it can be observed that the Ni 3$d$ states lie in the band gap of LaAlO$_{3}$, whereas the La 4$f$ states lie between 6 and 7 eV. The electronic properties in the vicinity of the Fermi level are governed by the Ni 3$d$ states admixed with O 2$p$ states, with negligible contribution from La and Al states, respectively. The system has a band gap of 0.92 eV and a  Ni$^{2+}$ spin magnetic moment of $1.44~\mu_\text{B}$.

The vibrational density of states (VDOS) in Fig.~\ref{Fig3} (right) exhibits no imaginary phonon frequencies, proving that the SL is dynamically stable. The La VDOS extends predominantly between 0-5~THz with the most prominent peak at 4~THz, constituting mainly the acoustic phonon modes along with a smaller contributions from oxygen. The O VDOS has four peaks between 5-20~THz. The Ni and Al VDOS are much lower, the former is more pronounced between 5-14 THz, while the latter between 7-17~THz. Notably, the Ni1 sites having half-filled $e_{g}$ orbitals have a VDOS peak between 5-10~THz, whereas the VDOS of the Ni2 ions with fully filled $t_{2g}$ orbitals shows four peaks with lower intensity between 5-20 THz. We further outline the essential role of the 1$\times$1$\times$2 supercell in achieving the lattice dynamical stability in the pristine SL. Using the smaller 1$\times$1$\times$1 supercell yielded finite imaginary phonon frequencies arising from the La atoms, shown in Fig. S4 (a) of the supplemental material~\cite{Appendix2020}. 

\begin{figure*}
\centering
\includegraphics[width=18.1cm, height=12.2cm]{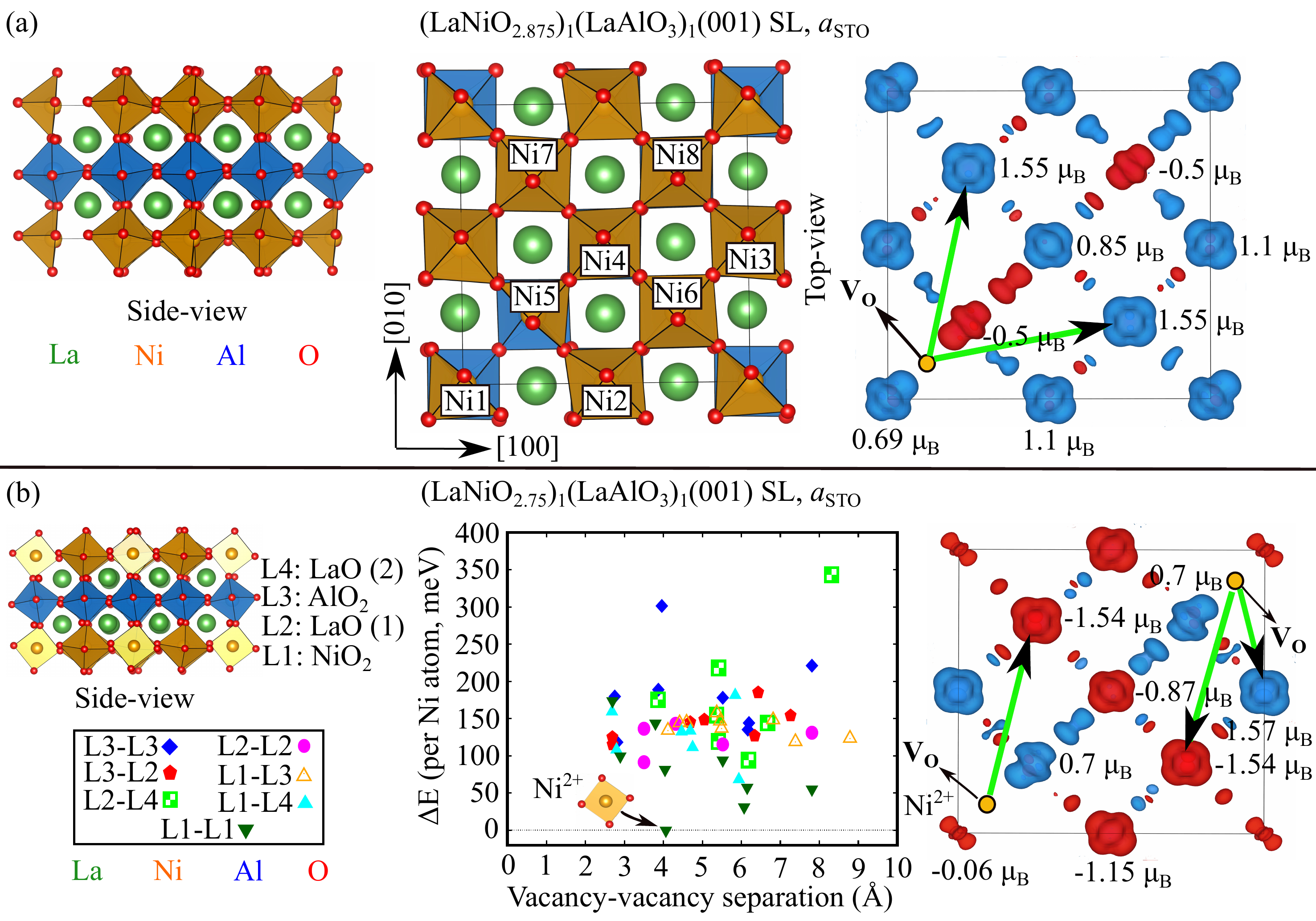}
\caption{\label{Fig:HS}(a) Side and top views of the optimized structure of (LaNiO$_{2.875}$)$_1$/(LaAlO$_{3}$)$_1$(001) SL and the corresponding spin-density plot at $a_{\mathrm{STO}}$. Ni sites with distinct spin moments and symmetry inequivalence are marked on the top view of the optimized SL structure. (b) Relative energy difference of all possible configurations for two oxygen vacancies in a (LaNiO$_{2.75}$)$_1$/(LaAlO$_{3}$)$_1$(001) SL at $a_{\mathrm{STO}}$. $\Delta E$ w.r.t. the most stable configuration as a function of the distance between the vacancies. The spin-density of the lowest energy case with Ni$^{2+}$ (LS) site in a fourfold coordination at $a_{\mathrm{STO}}$ is displayed on the right. L1 to L4 denote the planes in the SL. L$i$-L$j$ indicates in which layer the two vacancies are located. Blue and red colors denote positive and negative spin densities, respectively, integrated between -7~eV and the Fermi level. Arrows mark the oxygen vacancy sites and the localization sites for the released electrons from the oxygen vacancy are connected with green lines. The isovalue of the spin density is 0.006 e/\AA$^{3}$.}
\label{Fig4}
\end{figure*}

\section{\boldmath Electronic and magnetic properties of (L\lowercase{a}N\lowercase{i}O$_{3-\delta}$)$_1$/(L\lowercase{a}A\lowercase{l}O$_{3}$)$_1$(001) superlattices}\label{SEC6} 

Here we discuss the influence of oxygen vacancies on the electronic and magnetic properties of (LaNiO$_{3-\delta}$)$_1$/(LaAlO$_{3}$)$_1$(001) SL. The oxygen vacancy concentration ($\delta$) is varied from 0.125 to 0.5. The effect of strain is considered by setting the lateral lattice constant to bulk SrTiO$_3$ ($a_{\mathrm{STO}}$ = 3.905~{\AA}) and LaSrAlO$_4$ ($a_{\mathrm{LSAO}}$ = 3.756~{\AA}). The cross-plane lattice constants do not vary much with $\delta$: 3.75-3.77~{\AA} ($a_{\mathrm{STO}}$) and 3.86-3.88~{\AA} ($a_{\mathrm{LSAO}}$).

At $\delta$ = 0.125, we find that the oxygen vacancy (V$_{\mathrm{O}}$) prefers to occupy the NiO$_2$ plane (see Table I in section S-4-A of the supplemental material~\cite{Appendix2020}), irrespective of strain, similar to findings for lower oxygen vacancy concentrations~\cite{Geisler2022}. The orbital-projected DOS of all the Ni sites (see Fig.~\ref{Fig4} (a)) is displayed in Fig. S9 of the supplemental material~\cite{Appendix2020}), at both tensile and compressive strain, respectively. The localization of electrons released by the oxygen defects triggers a varying degree of CD throughout the LaNiO$_{3}$ sublattice, resulting in distinct spin moments at the Ni sites, shown in the spin density plot in Fig.~\ref{Fig4}(a) for $a_{\mathrm{STO}}$ (see Fig. S9 (b) of the supplemental material~\cite{Appendix2020} for $a_{\mathrm{LSAO}}$). With one oxygen vacancy (1/4 electron per Ni site), the two released electrons localize at distant Ni sites (Ni6 and Ni7) from the oxygen defect site rather than on an adjacent Ni site, connected by the green colored lines on the spin density plot in Fig.~\ref{Fig4} (a) for $a_{\mathrm{STO}}$ (see Fig. S9 (b) of the supplemental material~\cite{Appendix2020} for $a_{\mathrm{LSAO}}$). The orbital-projected DOS of Ni6 and Ni7 show that both the $e_g$ orbitals are occupied in the spin-up channel, irrespective of strain. This leads to enlarged octahedral volumes 10.76~{\AA$^3$} ($a_{\mathrm{STO}}$)/9.94~{\AA$^3$} ($a_{\mathrm{LSAO}}$) and a spin magnetic moment of 1.55~$\mu_\text{B}$ ($a_{\mathrm{STO}}$)/1.46~$\mu_\text{B}$ ($a_{\mathrm{LSAO}}$). Further sites with nonzero magnetic moments are Ni3 (1.1~$\mu_\text{B}$), Ni4 (0.85~$\mu_\text{B}$), Ni5 (-0.5~$\mu_\text{B}$), and Ni8 (-0.5~$\mu_\text{B}$),  the corresponding polyhedral volumes are 10~{\AA$^3$}, 9.5~{\AA$^3$}, 4.8~{\AA$^3$}, and 9.4~{\AA$^3$}, respectively. A similar variation in CD is also obtained at compressive strain  (see Fig. S9 (b) of the supplemental material~\cite{Appendix2020}). Overall, ferrimagnetic (FIM) order arises due to the negative spin moments of Ni5 and Ni8 polyhedra. This results in an enhanced bandwidth of the conduction band, especially in the spin-down channel and a band gap of 0.5 eV at $a_{\mathrm{STO}}$ and 0.32 eV at $a_{\mathrm{LSAO}}$, respectively. Apart from the FIM-CD phase, we find an insulating AFM phase, 26 meV/Ni atom at $a_{\mathrm{STO}}$ and 27.5 meV/Ni atom at $a_{\mathrm{LSAO}}$ less stable than FIM-CD. The LDOS of the two magnetic phases at tensile and compressive strains are provided in subsection S-4-B of the supplemental material~\cite{Appendix2020}. Furthermore, the relative stability of different possible magnetic orders is displayed in Table-II (for both tensile and compressive strain) and discussed in subsection S-4-A of the supplemental material~\cite{Appendix2020}. We find that one oxygen vacancy in the NiO$_{2}$ plane is 0.5 eV/SL more stable at $a_{\mathrm{LSAO}}$ than at $a_{\mathrm{STO}}$, similar to results at lower oxygen vacancy concentrations and at different compressive strain ($a_{\mathrm{LaAlO_{3}}}$=3.79~\AA)~\cite{Geisler2022}.  

\begin{figure}
\includegraphics[width=8.7cm, height=14.2cm]{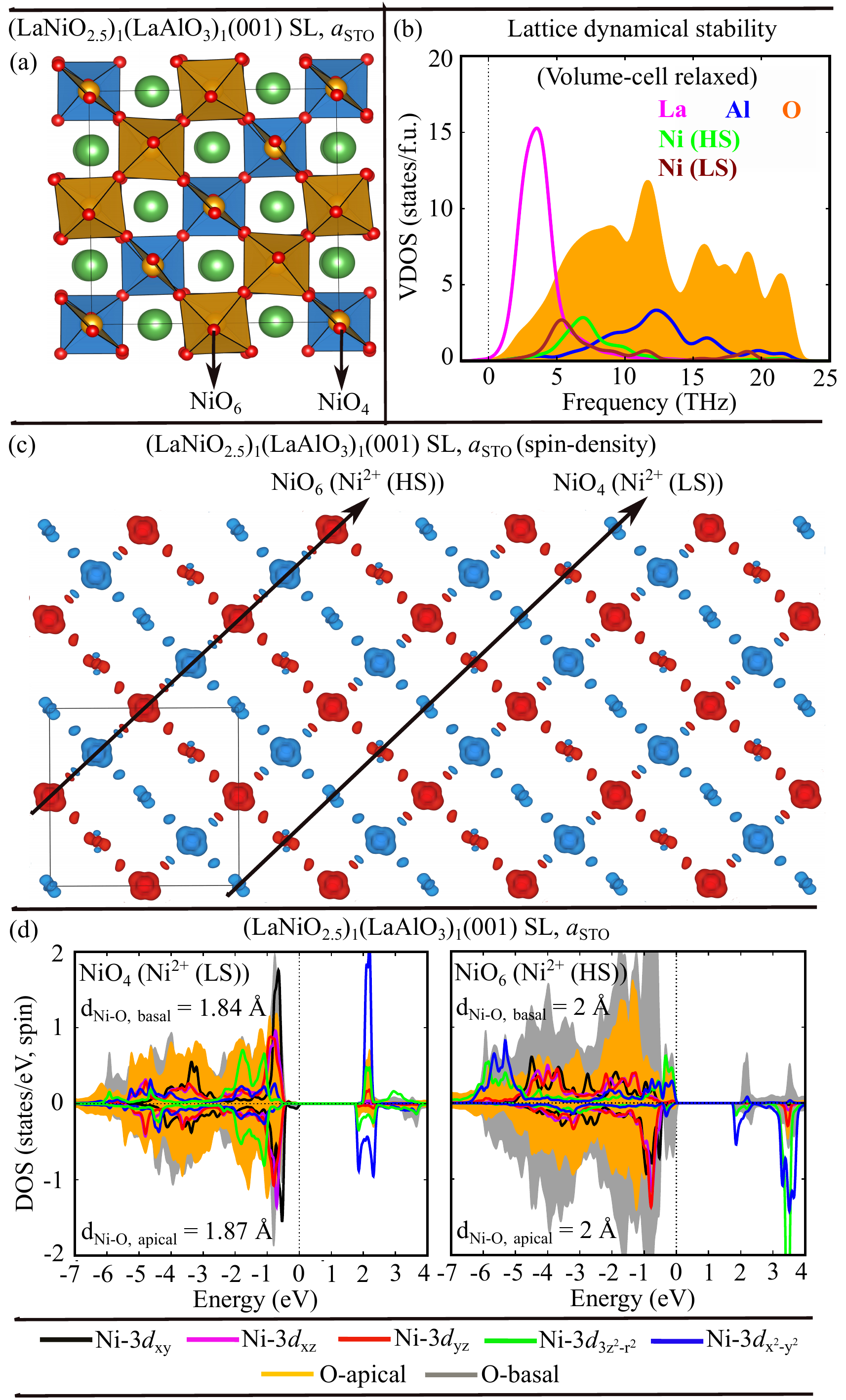}
\caption{\label{Fig:HS} (a) Top view of the optimized (LaNiO$_{2.5}$)$_1$/(LaAlO$_{3}$)$_1$(001) SL at $a_{\mathrm{STO}}$. (b) Atom-projected vibrational density of states of the volume-cell relaxed (LaNiO$_{2.5}$)$_1$/(LaAlO$_{3}$)$_1$(001) SL with stripe AFM magnetic order, obtained using $U=4$~eV on the Ni-3$d$ and  $U=7$~eV on the La-4$f$ states, respectively. In the vibrational density of states, magenta, green, brown, blue and orange colors represent La, Ni (HS), Ni (LS), Al, and O states, respectively. (c) Spin density integrated between -7~eV up to the Fermi level of the optimized (LaNiO$_{2.5}$)$_1$/(LaAlO$_{3}$)$_1$(001) SL at $a_{\mathrm{STO}}$ with isovalue 0.006 e/\AA$^{3}$. Blue and red colors correspond to positive and negative spin densities, respectively. (d) The orbital projected density of states of the Ni$^{2+}$ (LS), left and Ni$^{2+}$ (HS), right, respectively, along with the density of states of the apical and basal oxygen associated with the NiO$_{4}$ square planar plaquette (left) and the NiO$_{6}$ octahedron (right).}
\label{Fig5}
\end{figure}

To determine the energetically favorable positions for the oxygen di-vacancies at $\delta$ = 0.25 in the SLs, we  placed the two V$_{\mathrm{O}}$ at all possible sites and relaxed the internal coordinates, while keeping the volume fixed to the optimized pristine SL, obtained at $a_{\mathrm{STO}}$ and $a_{\mathrm{LSAO}}$, respectively. The relative total energies w.r.t. the most favorable configuration are displayed in Fig.~\ref{Fig4} (b) for $a_{\mathrm{STO}}$ (see Fig. S10 (a) of the supplemental material~\cite{Appendix2020} for $a_{\mathrm{LSAO}}$), we conclude that in the most favourable configuration the two oxygen vacancies are located along the [110] direction ($\approx$ 4.1~{\AA} at $a_{\mathrm{STO}}$ and $\approx$ 3.88~{\AA} at $a_{\mathrm{LSAO}}$) in the NiO$_2$ plane. With on average 1/2 electron per Ni site, 3 out of 4 released electrons localize at Ni3, Ni6 and Ni7 sites, respectively, resulting in spin  moments exceeding 1.5~$\mu_\text{B}$, irrespective of strain. The fourth electron occupies the spin-down $d_{3z^2-r^2}$ orbital at the square-planar NiO$_2$ site, leading to a quenched spin magnetic moment of -0.05$\mu_\text{B}$, and a band gap opening due to the crystal field splitting between the $e_g$ states, shown in the orbital-projected DOS (see Fig. S10 (b) for $a_{\mathrm{STO}}$ and Fig. S10 (c) for $a_{\mathrm{LSAO}}$ of the supplemental material~\cite{Appendix2020}). Similar to $\delta$ = 0.125, the localization of electrons released by oxygen defects in the NiO$_2$ plane leads to a FIM-CD phase, with three sites with  negative Ni spin moment (Ni2, Ni4, and Ni6), as shown in the spin density plot in Fig.~\ref{Fig4} (b) for $a_{\mathrm{STO}}$ (see Fig. S10 (a) of the supplemental material~\cite{Appendix2020} for $a_{\mathrm{LSAO}}$). This leads to an opening of a band gap of 0.6 eV at $a_{\mathrm{STO}}$ and 0.32 eV at $a_{\mathrm{LSAO}}$. Notably, the insulating FIM-CD phase, obtained in (LaNiO$_{2.75}$)$_1$/(LaAlO$_{3}$)$_1$(001) SL is distinct  from the zig-zag type AFM order with narrow band gap in LaNiO$_{2.75}$ bulk (DFT + $U$)~\cite{Shin2022}. Finally, all other possible initial configurations resulted in ferrimagnetic order, albeit less stable, at both tensile and compressive strains, respectively.

\begin{figure*}
\centering
\includegraphics[width=18.1cm, height=12cm]{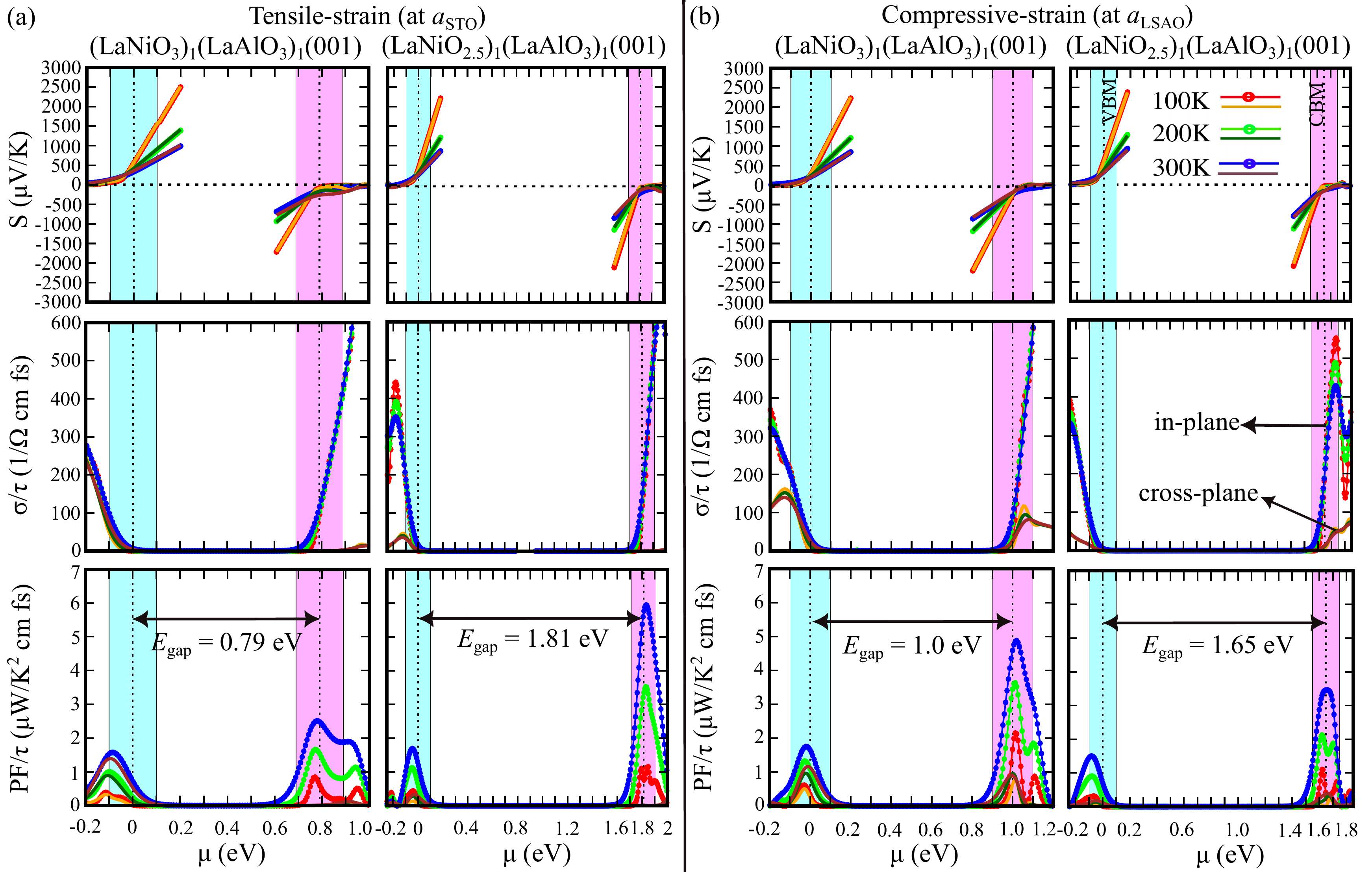}
\caption{\label{Fig:HS}Thermoelectric properties of (LaNiO$_{3-\delta}$)$_1$/(LaAlO$_{3}$)$_1$(001) SL ($\delta$ = 0 and 0.5) (a) at $a_{\mathrm{STO}}$ and (b) at $a_{\mathrm{LSAO}}$, for three different temperatures 100, 200, and 300~K. From top to bottom, Seebeck coefficient $S$, electrical conductivity $\sigma$/$\tau$, and the electronic power factor PF/$\tau$ are shown, where $\tau$ denotes the relaxation time. Orange, dark-green, and brown lines (red, green, and blue dotted lines) correspond to cross-plane (in-plane) transport. The vertical dashed lines denote the VBM and CBM. The energy range where the chemical potential is varied ($\pm$100 meV around VBM and CBM) is marked in cyan and magenta, respectively.}
\label{Fig6}
\end{figure*}

At $\delta$ = 0.5, guided by the $\delta$ = 0.25 case, we find that the four oxygen vacancies order along the [110] direction to form four NiO$_{4}$ plaquettes with square-planar geometry, as shown in Fig.~\ref{Fig5} (a). All other possible arrangements of the oxygen vacancies were found to be energetically less stable (see subsection S-4-A of the supplemental material~\cite{Appendix2020}), irrespective of strain. Out of the eight total released electrons, four electrons occupy the spin-down $d_{3z^2-r^2}$ orbital at Ni1, Ni4, Ni5, and Ni8 in the square-planar plaquettes, leading to Ni$^{2+}$ (LS), with a nominal magnetic moment of $\sim$ 0.1$\mu_\text{B}$, irrespective of strain, as shown in the orbital-projected DOS in Fig.~\ref{Fig5} for $a_{\mathrm{STO}}$ (see Fig. S11 (c) of the supplemental material~\cite{Appendix2020} for $a_{\mathrm{LSAO}}$). The remaining four electrons localize at Ni2, Ni3, Ni6, and Ni7, resulting in an increased spin magnetic moment of 1.63~$\mu_\text{B}$ ($a_{\mathrm{STO}}$) and 1.60~$\mu_\text{B}$ ($a_{\mathrm{LSAO}}$) due to the half-filled $e_{g}$ orbitals, corresponding to Ni$^{2+}$ (HS). Together they form stripes along the [110] direction  of alternating Ni$^{2+}$ (HS) in octahedral coordination and Ni$^{2+}$ (low-spin) in a fourfold planar configuration. This S-AFM order results in a significantly reduced Ni 3$d$ bandwidth and a larger band gap of 1.81 eV at $a_{\mathrm{STO}}$ and 1.65 eV at $a_{\mathrm{LSAO}}$ (see Fig. S12 for $a_{\mathrm{STO}}$ and Fig. S13 for $a_{\mathrm{LSAO}}$ of the supplemental material~\cite{Appendix2020}). Interestingly, the Ni spins couple antiferromagnetically along the Ni$^{2+}$ (HS) stripe, as shown in the spin density plot in Fig.~\ref{Fig5} (c). Notably, this is different from the A-type AFM order with a checkerboard ordering of NiO$_{4}$ and NiO$_{6}$ columns on the (001)$_{\mathrm{pc}}$ plane in bulk LaNiO$_{2.5}$~\cite{Shin2022}. The relative stability of different magnetic configurations is discussed in subsection S-4-A of the supplemental material~\cite{Appendix2020} for both tensile and compressive strain. From the above discussion, we conclude that the electron localization is the preferred mechanism to accommodate the excess electrons released from the oxygen vacancies, leading to MIT in (LaNiO$_{3-\delta}$)$_1$/(LaAlO$_{3}$)$_1$(001) SLs, irrespective of strain. We now turn to the lattice dynamic stability of (LaNiO$_{2.5}$)$_1$/(LaAlO$_{3}$)$_1$(001) SL in the S-AFM phase. Here also, we adopted the volume-cell relaxed structure. The element-resolved VDOS shown in Fig.~\ref{Fig5} (b) exhibits no imaginary phonon frequencies and  confirms the lattice dynamical stability of the SL. Similar to the pristine SL case, the acoustic phonon modes with a peak at $\sim$ 3 THz have La-character admixed with lower contribution from O VDOS between 0-5 THz. Notably the O VDOS peaks, lying between 0-23 THz differ from ones in the pristine SL, shown in Fig.~\ref{Fig3}. The Ni$^{2+}$ (HS) and Ni$^{2+}$ (LS) extend between 3-12 TH, the latter slightly shifted to lower frequences, while the Al VDOS lies between 5-20 THz. For the lattice dynamical stability it was essential to employ a 1$\times$1$\times$2 supercell (152 atoms). The 1$\times$1$\times$1 supercell containing 76 atoms  yielded instead  finite imaginary phonon frequencies arising from the La atoms, shown in Fig. S4 (b) of the supplement~\cite{Appendix2020}.

\begin{table}[t!]
	\caption{\label{Table-TE}Thermoelectric performance of the  (LaNiO$_{3-\delta}$)$_1$/(LaAlO$_{3}$)$_1$(001) SLs at different $\delta$ in comparison to a selection of prominent oxide thermoelectrics at room temperature (300~K).} 
	\begin{ruledtabular}
		\begin{tabular}{lcc}
			System & $S$ ($\mu$V/K) & PF ($\mu$W/K$^2$ cm)	\\
			\hline
			\hline
			\multicolumn{3}{c}{Tensile strain ($a_{\mathrm{STO}}$), $\tau = 4$~fs} \\
			\hline
			\hline
			(LaNiO$_{3}$)$_1$/(LaAlO$_{3}$)$_1$(001) & & \\
			in-plane (xx)& $-418$ & $10$  \\
			cross-plane (zz) & $-507$  & $0.4$  \\
			\hline
			(LaNiO$_{2.875}$)$_1$/(LaAlO$_{3}$)$_1$(001) & & \\
			in-plane (xx)& $-518$ & $9.4$  \\
			cross-plane (zz)& $-441$  & $0.6$  \\
			\hline
            (LaNiO$_{2.75}$)$_1$/(LaAlO$_{3}$)$_1$(001) & & \\
			in-plane (xx) & $-393$ & $0.2$  \\
			cross-plane (zz) & $-344$  & $0.3$  \\	
			\hline
            (LaNiO$_{2.5}$)$_1$/(LaAlO$_{3}$)$_1$(001) & & \\
			in-plane (xx) & $-467$ & $24$  \\
			cross-plane (zz) & $-399$ & $0.04$  \\
			\hline
			\hline
			\multicolumn{3}{c}{Compressive strain ($a_{\mathrm{LSAO}}$), $\tau = 4$~fs} \\
			\hline
			\hline
			(LaNiO$_{3}$)$_1$/(LaAlO$_{3}$)$_1$(00) & & \\
			in-plane (xx)& $-546$ & $19.5$  \\
			cross-plane (zz) & $-511$  & $3.7$  \\
			\hline
			(LaNiO$_{2.875}$)$_1$/(LaAlO$_{3}$)$_1$(001) & & \\
			in-plane (xx)& $-461$ & $4.5$  \\
			cross-plane (zz) & $-417$ & $0.3$  \\
			\hline
            (LaNiO$_{2.75}$)$_1$/(LaAlO$_{3}$)$_1$(001) & & \\
			in-plane (xx) & $-470$ & $0.6$  \\
			cross-plane (zz) & $-392$ & $0.2$  \\	
			\hline
            (LaNiO$_{2.5}$)$_1$/(LaAlO$_{3}$)$_1$(001) & & \\
			in-plane (xx) & $-400$ & $14$  \\
			cross-plane (zz) & $-428$  & $1.6$  \\
			\hline
			\hline			
			\multicolumn{3}{c}{Literature} \\
			\hline
			\hline
			SrTiO$_3$ (DFT\cite{Bilc2016}, $\tau$= 4.3 fs) & $-400$  & $10$\\    
			\hline	    
			La:SrTiO$_3$ bulk, exp.\cite{Okuda2001} & $-380$  & $35$  \\
			La:SrTiO$_3$ films, exp.\cite{Jalan2010} & $-980$  & $39$  \\
			Nb:SrTiO$_3$ bulk, exp.\cite{Ohta2005} & $-240$  & $20$  \\				
		\end{tabular}
	\end{ruledtabular}
\end{table}

\section{\boldmath Thermoelectric properties of (L\lowercase{a}N\lowercase{i}O$_{3-\delta}$)$_1$/(L\lowercase{a}A\lowercase{l}O$_{3}$)$_1$(001) superlattices} \label{SEC7}

Finally, in this section we discuss the thermoelectric properties of (LaNiO$_{3-\delta}$)$_1$/(LaAlO$_{3}$)$_1$(001) SLs at tensile and compressive strain, respectively, and compare them with other promising oxide thermoelectrics. We focus here on the $n$-type thermoelectric response. The thermoelectric properties are summarized in Fig.~\ref{Fig6} and Table~\ref{Table-TE}. We report power factors divided by the relaxation time PF$/\tau$ in Fig.~\ref{Fig6} which are hence independent of the choice of $\tau$ and related to the electronic fitness function introduced by Xing \textit{et al.}~\cite{Xing2017}. For Table~\ref{Table-TE}, we use a room-temperature relaxation time of $\tau = 4$~fs, a typical value for oxides~\cite{Bilc2016}. In our discussion we focus on a variation of the chemical potential ($\mu$) within a physically relevant $\pm 100$~meV interval around the band edges, highlighted with cyan  and magenta around the valence (VBM) and  conduction band minimum (CBM), respectively, for the calculation of the Seebeck coefficient. With the chemical potential shifted to the CBM we obtain an estimate for the $n$-type thermoelectric response at 300~K. In the pristine SL at $a_{\mathrm{STO}}$, the combination of flat and dispersive bands at the CBM leads to an in-plane electrical conductivity of 117 S cm$^{-1}$ fs$^{-1}$ (at $E_{\mathrm{F}}$), which together with the high Seebeck coefficient reaching up to -418 $\mu$V/K (in-plane) results in a power factor of 10~$\mu$W/K$^2$ cm (assuming a relaxation time $\tau = 4$~fs). On the other hand, at $a_{\mathrm{LSAO}}$, the bands at the CBM are relatively flat which leads to a reduced in-plane electrical conductivity of 90 S cm$^{-1}$ fs$^{-1}$, but a higher Seebeck coefficient reaching up to -546 $\mu$V/K (in-plane) giving rise to a higher power factor of 19.5~$\mu$W/K$^2$ cm than at $a_{\mathrm{STO}}$. 

For $\delta$ = 0.125 and especially for 0.25, the presence of localized flat bands at the CBM, results in appreciably high in-plane Seebeck coefficient, irrespective of strain, albeit lower than the corresponding values in the pristine SLs. However, the low values of the electrical conductivity (especially at $\delta$ = 0.25) lead to very low power factors (see Table~\ref{Table-TE} and Fig. S14 of the supplemental material~\cite{Appendix2020}). Interestingly, for $\delta$ = 0.5 at $a_{\mathrm{STO}}$, the combination of flat and dispersive bands in the vicinity of the CBM results in a high in-plane electrical conductivity of 131~S cm$^{-1}$ fs$^{-1}$ which together with a high Seebeck coefficient reaching up to -467 $\mu$V/K (in-plane) lead to robust power factor of 24~$\mu$W/K$^2$ cm (assuming a relaxation time $\tau = 4$~fs); almost twice compared to pristine SL at 300~K. In contrast, for $\delta$ = 0.5 at $a_{\mathrm{LSAO}}$, the bands in the vicinity of the CBM are more dispersive which results in a much higher in-plane electrical conductivity of 187 S cm$^{-1}$ fs$^{-1}$ but a reduced Seebeck coefficient of  up to -400 $\mu$V/K (in-plane). This leads to a lower power factor of 14~$\mu$W/K$^2$ cm than at $a_{\mathrm{STO}}$. Overall, the thermoelectric power-factor in (LaNiO$_{3-\delta}$)$_1$/(LaAlO$_{3}$)$_1$(001) SLs especially at tensile strain is higher than in the pristine SLs, which makes the oxygen-deficient SLs attractive for thermoelectric applications. Moreover, the introduction of a moderate oxygen-vacancy concentration turns out to be favorable for reducing the lattice thermal conductivity, as reported for SrTiO$_{3}$~\cite{Yu2008}. 

\section{\boldmath Summary}\label{SEC8}
In summary, using a combination of DFT + $U$ and Boltzmann transport theory calculations in the constant relaxation time approximation, we studied the electronic, magnetic and thermoelectric properties of (LaNiO$_{3-\delta}$)$_1$/(LaAlO$_{3}$)$_1$(001) SLs for $\delta$ = 0, 0.125, 0.25 and 0.5 under tensile ($a_{\mathrm{STO}}$) and compressive strain ($a_{\mathrm{LSAO}}$). In the pristine 1/1 SLs, we found that confinement leads to an antiferromagnetic charge-disproportionated (AFM-CD) ($d^{8}${$\underline L$}$^{2}$)$_{S=0}$($d^{8}$)$_{S=1}$ phase, irrespective of strain. At $\delta$ = 0.125 and 0.25 the localization of electrons released by oxygen defect(s) in the NiO$_2$ plane results in semiconducting ferrimagnetic charge-disproportionated (FIM-CD) phases at both tensile and compressive strain. Next at $\delta$ = 0.5, an insulating phase emerges with alternating stripes of octahedrally coordinated Ni$^{2+}$ (HS) and Ni$^{2+}$ (LS) in a square-planar geometry and oxygen vacancies ordered along the [110] direction (S-AFM), leading to a robust $n$-type in-plane power factor of 24~$\mu$W/K$^2$ cm at $a_{\mathrm{STO}}$ and 14~$\mu$W/K$^2$ cm at $a_{\mathrm{LSAO}}$, respectively, (assuming relaxation time $\tau = 4$~fs), at 300~K. In addition, the lattice dynamic stability was demonstrated for the pristine SL and for $\delta$ = 0.5. These findings suggest ways to design exotic phases using oxygen vacancies in ultrathin nickelate superlattices, that are potentially interesting for applications.

\begin{acknowledgments}
This work was supported by the German Research Foundation (Deutsche Forschungsgemeinschaft, DFG) within the SFB/TRR 80 (Projektnummer 107745057, subprojects G3 and G8). We acknowledge computational time at the Leibniz Rechenzentrum Garching, Project No. pr87ro and magnitUDE supercomputer (DFG Grants No. INST 20876/209-1 FUGG and No. INST 20876/243-1 FUGG).
\end{acknowledgments}

\end{document}